\documentstyle[prl , aps, multicol]{revtex}

\author{G.~Schmidt, D. ~Ferrand, L.~W.~Molenkamp}
\address{Physikalisches Institut, Universit\"at W\"urzburg, Am Hubland, 97074 W\"urzburg,
Germany}

\author{A.~T.~Filip, B.~J.~van Wees}

\address{Dept. of Applied Physics and Materials Science Centre, University of Groningen, \\ Nijenborgh
4, 9747 AG Groningen, the Netherlands }
\date{\today}
\title{A basic obstacle for electrical spin-injection from a ferromagnetic metal into a diffusive semiconductor}

\begin{document}

\maketitle

\begin{center}
(Accepted by PRB Rap. com.)
\end{center}

\begin{abstract}
We have calculated the spin-polarization effects of a current in a two dimensional electron gas
which is contacted by two ferromagnetic metals. In the purely diffusive regime, the current may
indeed be spin-polarized. However, for a typical device geometry the degree of spin-polarization of
the current is limited to less than 0.1\%, only. The change in device resistance for parallel and
antiparallel magnetization of the contacts is up to quadratically smaller, and will thus be
difficult to detect.
\end{abstract}

\begin{multicols}{2}

Spin-polarized electron injection into semiconductors has been a field of growing interest during
the last years\cite{Datta,Roukes,Aronov,Prinz}. The injection and detection of a spin-polarized
current in a semiconducting material could combine magnetic storage of information with electronic
readout in a single semiconductor device, yielding many obvious advantages. However, up to now
experiments for spin-injection from ferromagnetic metals into semiconductors have only shown
effects of less than 1\%\cite{Bland,Johnson}, which sometimes are difficult to separate from
stray-field-induced Hall- or magnetoresistance-effects\cite{Roukes}. In contrast, spin-injection
from magnetic semiconductors has already been demonstrated successfully\cite{Nature,Ohno} using an
optical detection method.

Typically, the experiments on spin-injection from a ferromagnetic contact are performed using a
device with a simple injector-detector geometry, where a ferromagnetic metal contact is used to
inject spin polarized carriers into a two dimensional electron gas (2DEG)\cite{Bland}. A
spin-polarization of the current is expected from the different conductivities resulting from the
different densities of states) for spin-up and spin-down electrons in the ferromagnet. For the full
device, this should result in a conductance which depends on the relative magnetization of the two
contacts\cite{Datta}.

A simple linear-response model for transport across a ferromagnetic/normal metal interface, which
nonetheless incorporates the detailed behaviour of the electrochemical potentials for both spin
directions was first introduced by van Son et al.\cite{VanSon}. Based on a more detailed
(Boltzmann) approach, the model was developed further by Valet and Fert for all metal multilayers
and GMR\cite{Valet}. Furthermore, it was applied by Jedema et al. to superconductor-ferromagnet
junctions\cite{Jedema}. For the interface between a ferromagnetic and a normal metal, van Son et
al. obtain a splitting of the electrochemical potentials for spinup and spindown electrons in the
region of the interface. The model was applied only to a single contact and its boundary
resistance\cite{VanSon}. We now apply a similar model to a system in which the material properties
differ considerably.

Our theory is based on the assumption that spin-scattering occurs on a much slower timescale than
other electron scattering events\cite{Oestreich}. Under this assumption, two electrochemical
potentials $\mu_\uparrow$ and $\mu_\downarrow$, which need not be equal, can be defined for both
spin directions at any point in the device\cite{VanSon}. If the current flow is one dimensional in
the $\bf x$-direction, the electrochemical potentials are connected to the current via the
conductivity $\sigma$, the diffusion constant D, and the spin-flip time constant $\tau_{sf}$ by
Ohm's law and the diffusion equation, as follows:

\begin{mathletters}
\label{j}

\begin{equation} \label{ja} \frac{\partial {\mu_{\uparrow,\downarrow}}}{\partial {x}}=-\frac{e\
j_{\uparrow,\downarrow}}{\sigma_{\uparrow,\downarrow}} \label{equationa}
\end{equation}
\begin{equation}
\frac{\mu_\uparrow-\mu_\downarrow}{\tau_{sf}}= \frac{D\ \partial^2{(\mu_\uparrow-
\mu_\downarrow)}}{\partial{x^2}}
\end{equation}
\end{mathletters}

where D is a weighted average of the different diffusion constants for both spin
directions\cite{VanSon}. Without loss of generality, we assume a perfect interface without spin
scattering or interface resistance, in a way that the electrochemical potentials
$\mu_{\uparrow\downarrow}$ and the current densities $j_{\uparrow\downarrow}$ are continuous.

Starting from these equations, straightforward algebra leads to a splitting of the electrochemical
potentials at the boundary of the two materials, which is proportional to the total current density
at the interface. The difference $(\mu_\uparrow-\mu_\downarrow)$ between the electrochemical
potentials decays exponentially inside the materials, approaching zero difference at $\pm \infty$.

\begin{equation}
\label{infinity}
(\mu_\uparrow(\pm\infty)=\mu_\downarrow(\pm\infty))
\end{equation}

A typical lengthscale for the decay of $(\mu_\uparrow-\mu_\downarrow)$ is the spin-flip length
$\lambda=\sqrt{D\tau_{sf}}$ of the material. In a semiconductor, the spin-flip length
$\lambda_{sc}$ can exceed its ferromagnetic counterpart $\lambda_{fm}$ by several orders of
magnitude. In the limit of infinite $\lambda_{sc}$, this leads to a splitting of the
electrochemical potentials at the interface which stays constant throughout the semiconductor. If
the semiconductor extends to $\infty$, Eqs. \ref{j} in combination with Eq. \ref{infinity} imply a
linear and parallel slope of the electrochemical potentials for spin-up and spin-down in the
semiconductor, forbidding injection of a spin-polarized current if the conductivities for both spin
channels in the 2DEG are equal. At the same time, we see that the ferromagnetic contact influences
the electron system of the semiconductor over a lengthscale of the order of the spin-flip length in
the semiconductor. A second ferromagnetic contact applied at a distance smaller than the spin-flip
length may thus lead to a considerably different behaviour depending on its spin-polarization.

In the following, we will apply the theory to a one dimensional system in which a ferromagnet
(index $i=1$) extending from $x=-\infty$ to $x=0$ is in contact with a semiconductor (index $i=2$,
$0<x<x_0$), which again is in contact to a second ferromagnet (index $i=3$, $x_0\leq x\leq
\infty$). This system corresponds to a network of resistors $R_{1\uparrow,\downarrow}$,
$R_{SC\uparrow,\downarrow}$, and $R_{3\uparrow,\downarrow}$, representing the two independent spin
channels in the three different regions as sketched in Fig. (1a).

The (x dependent) spin-polarization of the current density at position $x$ is defined as
\begin{equation}
\alpha_i(x)\equiv\frac{j_{i\uparrow}(x)-j_{i\downarrow}(x)}{j_{i\uparrow}(x)+j_{i\downarrow}(x)}
\end{equation}

where we set the bulk spin polarization in the ferromagnets far from the interface
$\alpha_{1,3}(\pm\infty)\equiv\beta_{1,3}$. The conductivities for the spin-up and spin-down
channels in the ferromagnets can now be written as
$\sigma_{1,3\uparrow}=\sigma_{1,3}(1+\beta_{1,3})/2$ and
$\sigma_{1,3\downarrow}=\sigma_{1,3}(1-\beta_{1,3})/2$. We assume that the physical properties of
both ferromagnets are equal, but allow their magnetization to be either parallel ($\beta_1=\beta_3$
and $R_{1\uparrow,\downarrow}=R_{3\uparrow,\downarrow}$) or antiparallel ($\beta_1=-\beta_3$ and
$R_{1\uparrow,\downarrow}=R_{3\downarrow,\uparrow}$). In the linear-response regime, the difference
in conductivity for the spin-up and the spin-down channel in the ferromagnets can easily be deduced
from the Einstein relation with $D_{i\uparrow}\neq D_{i\downarrow}$ \cite{Jedema} and
$\rho_{i\uparrow}(E_F)\neq\rho_{i\uparrow}(E_F)$, where $\rho(E_F)$ is the density of states at the
Fermi energy, and D the diffusion constant.

To separate the spin-polarization effects from the normal current flow, we now write the
electrochemical potentials in the ferromagnets for both spin directions as
$\mu_{\uparrow,\downarrow}=\mu^0+\mu^*_{\uparrow,\downarrow}$, $(i=1,3)$, $\mu^0$ being the
electrochemical potential without spin effects. For each part $i$ of the device, Eqs. (\ref{j})
apply separately.

As solutions for the diffusion equation, we make the Ansatz

\begin{equation}
\label{solution} \mu_{i \uparrow,\downarrow}= \mu_i^0+\mu^*_{i \uparrow,\downarrow}= \mu_i^0+
c_{i\uparrow,\downarrow}\ \exp{\pm((x-x_i)/\lambda_{fm})}
\end{equation}
for $i=1,3$ with $x_1=0$, $x_3=x_0$, and the + (-) sign referring to index $1$ ($3$), respectively.

From the boundary conditions $\mu_{1\uparrow}(-\infty)=\mu_{1\downarrow}(-\infty)$ and
$\mu_{3\uparrow}(\infty)=\mu_{3\downarrow}(\infty)$, we have that the slope of $\mu^0$ is identical
for both spin directions, and also equal in region 1 and 3 if the conductivity $\sigma$ is
identical in both regions, as assumed above. In addition, these boundary conditions imply that the
exponential part of $\mu$ must behave as $c\  \exp{(x/\lambda_{fm})}$ in region 1 and as $c\
\exp{(-(x-x_0)/\lambda_{fm})}$ in region 3.

In the semiconductor we set $\tau_{sf}=\infty$, based on the assumption that the spin-flip length
$\lambda_{sc}$ is several orders of magnitude longer than in the ferromagnet and much larger than
the spacing between the two contacts. This is correct for several material systems, as
semiconductor spin-flip lengths up to $100 \mu$m have already been demonstrated \cite{Avshalom}. In
this limit, we thus can write the electrochemical potentials for spin-up and spin-down in the
semiconductor as

\begin{equation}
\mu_{2\uparrow,\downarrow}(x)=\mu_{1\uparrow,\downarrow}(0)+\gamma_{\uparrow,\downarrow}\  x \mbox
, \qquad \gamma_{\uparrow,\downarrow}=constant
\end{equation}

While the conductivities of both spin-channels in the ferromagnet are different, they have to be
equal in the two dimensional electron gas. This is because in the 2DEG, the density of states at
the Fermi level is constant, and in the diffusive regime the conductivity is proportional to the
density of states at the Fermi-energy. Each spin channel will thus exhibit half the total
conductivity of the semiconductor ($\sigma_{2\uparrow,\downarrow}=\sigma_{sc}/2$).

If we combine equation \ref{j} and \ref{solution} and solve in region 1 at the boundary $x=0$ and
in region 3 at $x=x_0$ we are in a position to sketch the band bending in the overall device. From
symmetry considerations and the fact that $j_{2\uparrow}$ and $j_{2\downarrow}$ remain constant
through the semiconductor (no spin-flip) we have

\begin{equation}
\mu_{1\uparrow}(0)-\mu_{1\downarrow}(0)=\pm(\mu_{3\downarrow}(x_0)-\mu_{3\uparrow}(x_0))
\end{equation}

where the +(-) sign refers to parallel (antiparallel) magnetization, respectively. This yields
$c_{1\uparrow}=-c_{3\uparrow}$ and $c_{1\downarrow}=-c_{3\downarrow}$ in the expression for
$\mu_{\uparrow\downarrow}$ in Eq. \ref{solution}  for the parallel case, which is shown
schematically in Fig. (1b). The antisymmetric splitting of the electrochemical potentials at the
interfaces leads to a different slope and a crossing of the electrochemical potentials at
$x=x_0/2$. We thus obtain a different voltage drop for the two spin directions over the
semiconductor, which leads to a spin polarization of the current. In the antiparallel case where
the minority spins on the left couple to the majority spins on the right the solution is
$c_{1\uparrow}=-c_{3\downarrow}$ and $c_{1\downarrow}=-c_{3\uparrow}$ with
$j_{\uparrow}=j_{\downarrow}$. A schematic drawing is shown in Fig. (1c). The splitting is
symmetric and the current is unpolarized.

The physics of this result may readily be understood from the resistor model (Fig. 1a). For
parallel (antiparallel) magnetization we have $R_{1\uparrow}+R_{3\uparrow}\neq
R_{1\downarrow}+R_{3\downarrow}$ $(R_{1\uparrow}+R_{3\uparrow}=R_{1\downarrow}+R_{3\downarrow})$,
respectively. Since the voltage across the complete device is identical for both spin channels,
this results either in a different (parallel) or an identical (antiparallel) voltage drop over
$R_{SC\uparrow}$ and $R_{SC\downarrow}$.

For parallel magnetization ($\beta_1=\beta_3=\beta$) the finite spin-polarization of the current
density in the semiconductor can be calculated explicitly by using the continuity of
$j_{i\uparrow,\downarrow}$ at the interfaces under the boundary condition of charge conservation
for $(j_{i\uparrow}+j_{i\downarrow})$ and may be expressed as:

\begin{equation}
\label{beta} \alpha_2=\beta\mbox\ \frac{\lambda_{fm}}{\sigma_{fm}}\mbox\
\frac{\sigma_{sc}}{x_0}\mbox\ \frac{2}{(2{\frac{\lambda_{fm}\sigma_{sc}}{x_0\sigma_{fm}}}
+1)-\beta^2}
\end{equation}

where $\alpha_2$ is evaluated at $x=0$ and constant throughout the semiconductor, because above we
have set $\tau_{sf}=\infty$ in the semiconductor.

For a typical ferromagnet, $\alpha_2$ is dominated by
$(\lambda_{fm}/\sigma_{fm})/(x_0/\sigma_{sc})$ where $x_0/\sigma_{sc}$  and
$\lambda_{fm}/\sigma_{fm}$ are the resistance of the semiconductor and the relevant part of the
resistance of the ferromagnet, respectively. The maximum obtainable value for $\alpha_2$ is
$\beta$.

However, this maximum can only be obtained in certain limiting cases, i.e., $x_0\rightarrow 0$,
$\sigma_{sc}/\sigma_{fm}\rightarrow \infty$, or $\lambda_{fm}\rightarrow\infty$, which are far away
from a real-life situation. If, e.g., we insert some typical values for a spin injection device
($\beta=60\%$, $x_0=1 \mu$m, $\lambda_{fm}=10$ nm, and $\sigma_{fm}=10^4\sigma_{sc}$), we obtain
$\alpha\approx0.002\%$. The dependence of $\alpha_2$ on the various parameters is shown graphically
in Figs. (2a) and (2b) where $\alpha_2$ is plotted over $x_0$ and $\lambda_{fm}$, respectively, for
three different values of $\beta$. Apparently, even for $\beta>80\%$, $\lambda_{fm}$ must be larger
than $100$ nm or $x_0$ well below $10$ nm in order to obtain significant (i.e. $>1\%$) current
polarization. The dependence of $\alpha_2$ on $\beta$ is shown in Fig. (3a) for three different
ratios $\sigma_{fm}/\sigma_{sc}$. Even for a ratio of 10, $\alpha_2$ is smaller than $1\%$ for
$\beta<98\%$, where the other parameters correspond to a realistic device.

By calculating the electrochemical potential throughout the device we may also obtain $R_{par}$ and
$R_{anti}$ which we define as the total resistance in the parallel or antiparallel configuration,
respectively. The resistance is calculated for a device with ferromagnetic contacts of the
thickness $\lambda_{fm}$, because only this is the lengthscale on which spin dependent resistance
changes will occur. In a typical experimental setup, the difference in resistance $\Delta
R=(R_{anti}-R_{par})$ between the antiparallel and the parallel configuration will be measured. To
estimate the magnitude of the magnetoresistance effect, we calculate $\Delta R/R_{par}$ and we
readily find

\begin{equation}
\frac{\Delta R}{R_{par}}=\frac{\beta^2}{1-\beta^2}\mbox\ \frac{\lambda_{fm}^2}{\sigma_{fm}^2}\mbox\
\frac{\sigma_{sc}^2}{x_0^2}\mbox\ \frac{4}{(2{\frac{\lambda_{fm}\sigma_{sc}}{x_0\sigma_{fm}}}
+1)^2-\beta^2}
\end{equation}

Now, for metallic ferromagnets, $\Delta R/R_{par}$ is dominated by
$(\lambda_{fm}/\sigma_{fm})^2/(x_0/\sigma_{sc})^2$ and is approx. $\alpha_2^2$. In the limit of
$x_0\rightarrow 0$, $\sigma_{sc}/\sigma_{fm}\rightarrow \infty$, or
$\lambda_{fm}\rightarrow\infty$, we again obtain a maximum which is now given by

\begin{equation}
\label{deltar} \frac{\Delta R}{R_{par}}=\frac{\beta^2}{(\beta-1)(\beta+1)}
\end{equation}
Fig (3b) shows the dependence of $\alpha_2$ and $\Delta R/R_{par}$ on $\beta$, for a realistic set
of parameters. Obviously, the change in resistance will be difficult to detect in a standard
experimental setup.

We have thus shown, that, in the diffusive transport regime, for typical ferromagnets only a
current with small spin-polarization can be injected into a semiconductor 2DEG with long spin-flip
length even if the conductivities of semiconductor and ferromagnet are equal (Fig (3a)). This
situation is dramatically exacerbated when ferromagnetic metals are used; in this case the
spin-polarization in the semiconductor is negligible.

Evidently, for efficient spin-injection one needs a contact where the spin-polarization is almost
100\%. One example of such a contact has already been demonstrated: the giant Zeeman-splitting in a
semimagnetic semiconductor can be utilized to force all current-carrying electrons to align their
spin to the lower Zeeman level\cite{Nature}. Other promising routes are ferromagnetic
semiconductors\cite{Ohno} or the so called Heusler compounds\cite{deGroot} or other half-metallic
ferromagnets\cite{Park,Guentherodt}. Experiments in the ballistic transport regime\cite{Roukesbal}
(where $\sigma_{sc}$ has to be replaced by the Sharvin contact resistance) may circumvent part of
the problem outlined above. However, a splitting of the electrochemical potentials in the
ferromagnets, necessary to obtain spin-injection, will again only be possible if the resistance of
the ferromagnet is of comparable magnitude to the contact resistance.

\acknowledgments This work was supported by the European Commission (ESPRIT-MELARI consortium
'SPIDER'), the German BMBF under grant \#13N7313 and the Dutch Foundation for Fundamental Research
FOM.

\end{multicols}

\begin{figure}
\caption{(a) Simplified resistor model for a device consisting of a semiconductor (SC) with two
ferromagnetic contacts (FM) 1 and 3. The two independent spin channels are represented by the
resistors  $R_{1\uparrow,\downarrow}$, $R_{SC\uparrow,\downarrow}$, and $R_{3\uparrow,\downarrow}$.
(b) and (c) show the electrochemical potentials in the three different regions for parallel (b) and
antiparallel (c) magnetization of the ferromagnets. The solid lines show the potentials for spin-up
and spin-down electrons, the dotted line for $\mu_0$ (undisturbed case). For parallel magnetization
(b), the slopes of the electrochemical potentials in the semiconductor are different for both spin
orientations. They cross in the middle between the contacts. Because the conductivity of both spin
channels is equal, this results in a (small) spin-polarization of the current in the semiconductor.
In the antiparallel case (c), the slopes of the electrochemical potentials in the semiconductor are
equal for both spin orientations, resulting in unpolarized current flow. (Note that the slope of
$\mu$ in the metals is exaggerated).}
\end{figure}

\begin{figure}
\caption{Dependence of $\alpha_2$ on $\lambda_{fm}$ (a) and $x_0$ (b), respectively for
$\sigma_{fm}=100\ \sigma_{sc}$ and three different values of $\beta$. In fig. (a), $x_0$ is $1\mu
m$. Note that $\alpha_2$ is only in the range of \% for $\beta\approx100\%$ or $\lambda_{fm}$ in
the $\mu m$-range. In fig. (b) we have $\lambda_{fm}=10$ nm and again, we see that for a contact
spacing of more than $10$ nm, $\alpha_2$ will be below $1\%$ if a standard ferromagnetic metal
($\beta<80\%$) is used.}
\end{figure}

\begin{figure}
\caption{Dependence of $\alpha_2$ and $\Delta R/R$ on $\beta$. In (a) $\alpha_2$  is plotted over
$\beta$ for different ratios $\sigma_{fm}/\sigma_{sc}$. For a ratio of 100, $\alpha_2$ is well
below $0.1\%$ for $\beta<99\%$. In (b), again $\alpha_2$ is plotted versus $\beta$ with
$\sigma_{fm}/\sigma_{sc}=100$, with the corresponding values for $\Delta R/R$ on a logarithmic
scale. For $\beta$ between 0 and $90\%$, $\Delta R/R$ is smaller than $10^{-7}$ and thus difficult
to detect in the experiment.}
\end{figure}


\begin{references}
\bibitem{Datta}S. Datta, B. Das, Appl.\ Phys.\ Lett.\ {\bf56} (7), 665 (1990).
\bibitem{Roukes}F. G. Monzon, M. Johnson, M. L. Roukes, Appl.\ Phys.\ Lett.\ {\bf71} (21), 3087 (1997).
\bibitem{Aronov}A. G. Aronov, G. E. Pikus, Sov.\ Phys.\ Semicond.\ {\bf10} (6), 698 (1976).
{\bf15} (3), 1215, (1997)
\bibitem{Prinz}G. A. Prinz, Physics Today, {\bf48}(4), 58-63, (1995)
\bibitem{Bland}W. Y. Lee, S. Gardelis, B. C. Choi, Y. B. Xu, C. G.
Smith, C. H. W. Barnes, D. A. Ritchie, E. H. Linfield, J. A. C. Bland, J.\ Appl.\ Phys.\ {\bf85}(9), 6682 (1999)
\bibitem{Johnson}P. R. Hammar, B. R. Bennet,M. J. Yang, M. Johnson, Phys.\ Rev.\ Lett.\ {\bf83} (1), 203-206, (1999)
\bibitem{Nature}R. Fiederling, M. Keim, G. Reuscher, W. Ossau, G. Schmidt, A. Waag, L. W.
Molenkamp, Nature, 402, (1999), 787
\bibitem{Ohno} Y. Ohno, D. K. Young, B. Beschoten, F. Matsukura, H. Ohno, D. D. Awschalom, Nature, 402, (1999), 790
\bibitem{VanSon}P. C. van Son, H. van Kempen, P. Wyder, Phys.\
Rev.\ Lett.\ {\bf58}(21), 2271 (1987)
\bibitem{Valet}T. Valet, A. Fert, Phys.\ Rev.\ B\ {\bf48} 10, 7099 (1993)
\bibitem{Jedema}F. J. Jedema, B. J. van Wees, B. H. Hoving, A. T. Filip, T. M. Klapwijk,
Phys.\ Rev.\ B\ {\bf60}, 16549 (1999)
\bibitem{Oestreich}D. H\"agele, M. Oestreich, W. W. R\"uhle, N. Nestle, K. Eberl, Appl.\ Phys.\ Lett.\ {\bf73}
(11), 1580, (1998)
\bibitem{Avshalom} J. M. Kikkawa, D. D. Awschalom, Nature {\bf397}, 139-141, (1999)
\bibitem{deGroot}R. A. de Groot, F. M. M\"uller, P. G. van Engen, K. H. J. Buschow, Phys.\ Rev.\
Lett.\ {\bf50}, 2024 (1983)
\bibitem{Park}J.-H. Park, E. Vescovo, H.-J.Kim, C. Kwon, R. Ramesh, T. Venkatesan, Nature
{\bf392}, 794 (1998)
\bibitem{Guentherodt}K. P. K\"amper, W. Schmidt, G. G\"untherodt, R. J. Gambin, R. Ruf, Phys.\
Rev.\ Lett.\ {\bf59}, 2788 (1988)
\bibitem{Roukesbal} H. X. Tang, F. G. Monzon, R. Lifshitz, M. C.
Cross, M. L. Roukes, Phys.\ Rev.\ B\ {\bf61} 7 , 4437 (1999)




\end{references}
\end{document}